# A review and evaluation of standard methods to handle missing data on time-varying confounders in marginal structural models

Clémence Leyrat[1*], James R. Carpenter[1,2], Sébastien Bailly[3], Elizabeth J. Williamson[1,4]

[1] London School of Hygiene and Tropical Medicine, Department of Medical Statistics, London, UK
[2] London Hub for Trials Methodology Research, MRC Clinical Trials Unit, UCL, London, UK
[3] HP2, Inserm 1042, Université Grenoble Alpes, Grenoble, France
[4] Health Data Research London site, UK

*Corresponding author:
Dr Clémence Leyrat
Department in Medical Statistics
London School of Hygiene and Tropical Medicine
Keppel Street, WC1E 7HT London, UK
+44 (0)207 927 2169
clemence.leyrat@lshtm.ac.uk

**Funding:** This work was supported by the Medical Research Council (Project Grant MR/M013278/1).

## Abstract

*Background:* Marginal structural models (MSMs) are commonly used to estimate causal intervention effects in longitudinal non-randomised studies. A common issue when analysing data from observational studies is the presence of incomplete confounder data, which might lead to bias in the intervention effect estimates if they are not handled properly in the statistical analysis. However, there is currently no recommendation on how to address missing data on covariates in MSMs under a variety of missingness mechanisms encountered in practice.

*Methods:* We reviewed existing methods to handling missing data in MSMs and performed a simulation study to compare the performance of complete case (CC) analysis, the last observation carried forward (LOCF), the missingness pattern approach (MPA), multiple imputation (MI) and inverse-probability-of-missingness weighting (IPMW). We generated data in which the covariates and treatment were measured at three time-points and a continuous outcome was measured at the end of follow-up. Two covariates were partially observed. We considered three mechanisms for non-monotone missing data which are common in observational studies using electronic health record data. We illustrated the strengths and limitations of these analysis methods through an application to observational data looking at the causal effect of observance to continuous positive airway pressure device on sleepiness symptoms among patients with sleep apnoea.

*Results:* Whereas CC analysis lead to biased estimates of the intervention effect in almost all scenarios, the performance of the other approaches varied across scenarios. The LOCF approach led to unbiased estimates only under a specific non-random mechanism in which confounder values were missing when their values remained unchanged since the previous measurement. In this scenario, MI, the MPA and IPMW were biased. MI and IPMW led to the estimation of unbiased effects when data were missing at random, given the covariates or the treatment but only MI was unbiased when the outcome was a predictor of missingness. Furthermore, IPMW generally lead to very large standard errors. Lastly, regardless of the missingness mechanism, the MPA led to unbiased estimates only when the failure to record a confounder at a given time-point modified the subsequent relationships between the partially observed covariate and the outcome.

*Conclusions:* The choice of the appropriate method(s) to handle partially observed confounders when using MSMs must rely on a careful consideration of the reasons for missingness and whether missingness modifies the existing relationships among observed data. This choice also depends on the scientific context and the data source. Sensitivity analyses should be performed to assess the robustness of the results to departures from the postulated missingness mechanisms.

**Keywords:** missing data; marginal structural models; time-varying confounding; propensity score; multiple imputation; inverse probability weighting; missingness pattern approach; last observation carried forward; complete cases.

## 1. Introduction

Although randomised trials are often considered as the gold standard to establish causal effects of treatments and non-pharmacological interventions on health outcomes, the use of observational data for causal inference has been continuously increasing in the past 30 years (1). The enormous potential offered by the wealth of routinely collected medical data already available and the need for real-world evidence to assess the efficacy and safety of treatments have contributed to this phenomenon.

These routinely collected data have typically a longitudinal structure, as individuals are followed up over time, which allows the measurement of dynamic treatment patterns, including treatment switching or delay to treatment initiation. For instance, patients with chronic conditions often have a non-linear treatment history: treatment prescription might be updated based on the occurrence of new health events, changes in individual factors or side-effects induced by previous treatments. The newly prescribed treatment might, in turn, affect future health events and individual factors, themselves potentially associated with the outcome of interest (2). In such settings, specific statistical methods are required to account for confounding bias induced by time-varying variables (3).

In epidemiology, multivariable regression (4) and propensity scores (5,6) are often used to address confounding. When neither treatment nor confounders are time-varying multivariable regression can appropriately account for confounding and lead to an unbiased estimate of the (conditional) causal effect of the treatment. However, with time-varying treatment and confounders, adjusting for the confounders and treatment history is not sufficient, and often leads to biased estimates of the causal treatment effects (7). This is because the effect of a treatment received at a specific time on the outcome is mediated by subsequent treatments. On the other hand, propensity scores - defined as the individual probabilities of receiving the treatment of interest conditionally on individual characteristics - have been extended to situations with time-varying treatment and confounders (8), with scores estimated at each time-point (9). The cumulative product of the inverse of these scores over time can be used as a weight to account for confounding in the estimation of the treatment effect in a marginal structural model (MSM) (10). This method is by far the most common method to adjust for time-varying confounders in practice, followed by g-computation (11).

However, a common challenge in the statistical analysis of observational data is incomplete confounder information. In routinely collected data (e.g. from primary care), missing data may occur on treatments and outcomes but is particularly prevalent on covariates. This can happen if some information is not recorded at a given time-point or when the frequency of the measurement varies from one patient to another (e.g. asynchronous visits at the general practice (12)). The risk of missing confounder values usually increases with the number of time-points. This might jeopardize the validity of the results if the issue is ignored in the analysis, depending on the underlying missingness mechanisms. In practice, despite the STROBE recommendations to report the amount of missing data and the way they are handled (13) in observational studies, reporting is often suboptimal. A review looking at the reporting of missing exposure data in a longitudinal cohort showed that 43% of identified publications adhered to these guidelines (14). More importantly, when the method for handling missing data was reported, it was often done using inadequate methods. Although a variety of methods to handle missing data on covariates have been used in the context of time-varying exposures (14), the most common approaches – complete case (CC) analysis and last observation carried forward (LOCF) – have been criticized. The use of more complex approaches such as multiple imputation (MI) or inverse-probability of-missingness-weighting (IPMW) have been suggested, but the their performance is yet to be explored in a wider range of plausible missingness mechanisms. Another promising approach, which has not been extended to the context of MSMs, is the missingness pattern approach (MPA).

To our knowledge, there are no published guidelines for the choice of methods to handle missing confounder data in MSMs, and approaches used in simpler settings have not been investigated jointly with the use of MSMs. Published studies on missing data in MSMs focused on missing data in the exposure (14,15) or compared the performances of a few methods only (16,17). Moodie et al. (17) compared the use of IMPW and MI, showing that MI systematically outperformed IPMW, but they did not investigate the relative performance of the MPA and LOCF. Moreover, only one covariate and two time-points were generated in their simulation study, limiting the generalisability of the results. Vourli and Touloumi (18) investigated the performance of MI, IPMW and LOCF but found opposite conclusions in their setting, and showed that IPMW usually performed better than MI. This result might be explained by the omission of the outcome from the imputation model. A recent plasmode simulation (16) suggested superiority of MI over IPMW, but surprisingly, in their settings, CC analysis was the least biased. Furthermore, missingness mechanisms commonly studied in the literature (18) are those classically described under Rubin's taxonomy of missing data (19), but this taxonomy may be too restrictive to describe complex missingness scenarios encountered in routinely collected data (20). Therefore, the aim of this paper is to provide an overview of existing methods followed with practical guidelines to handle missing data on confounders in MSMs. These guidelines will rely on the understanding of the assumptions and missingness mechanisms under which these methods are valid. We will also present a simulation study comparing the performance of CC analysis, LOCF, MI, IPMW and MPA to handle partially observed confounders under three common missingness mechanisms encountered in observational studies.

The paper is organised as follows: Section 2 describes the use of MSMs. The issue of missing data and the available methods are discussed in Section 3. Section 4 and 5 present the design and the results of a simulation study, aiming to empirically evaluate the performance of CC analysis, LOCF, MI, IPMW and MPA to estimate the causal effect of a time-varying binary treatment on a continuous outcome. An illustrative example looking at the impact of the observance to treatment on sleepiness in patients with sleep apnoea is presented in Section 6, followed by a discussion (Section 7).

## 2. Causal inference in presence of time-varying treatment and confounders

In the presence of time-varying confounding, standard regression approaches fail because of treatment-confounder feedback (21), even when past treatment and confounders values are adjusted for (3). Robins (10) developed a new class of models (MSMs) to estimate causal effects in this setting. MSMs rely on an extension of inverse-probability-of-treatment weighting (IPTW) for more than one time-point within the potential outcomes framework. Details about this framework and the underlying assumptions for a single time-point can be found in Appendix A1. Similar to propensity score approaches, MSMs are a two-stage process. In the first stage, (21), weights – based on the inverse of the probability of a patient receiving the treatment they actually received – are estimated to create a pseudo-population in which the treatment is independent of the confounders. In the second stage, a weighted regression (using the weights derived in the first stage) including only the treatment history can be used to obtain estimate the causal effect of the treatment regimens of interest. Under the assumptions of no interference, consistency, exchangeability and positivity extended to time-varying settings and assuming the model used to obtain the weights is correctly specified, MSMs lead to unbiased estimates of the marginal causal effect of the treatment regimen.

In practice, the weights can be estimated using pooled logistic regression (9), in which each person-time interval is considered as an observation. This pooled logistic regression model must include the confounders and their relevant interactions to ensure the distributions of confounders are balanced between treatment groups in the weighted pseudo-population at each time-point. Further details about the implementation of MSMs can be found in Appendix A2.

## 3. Missing data in MSMs

### 3.1. Missingness patterns and missingness mechanisms in longitudinal studies

The choice of an appropriate procedure for addressing missing data relies on the characterization of the missingness patterns in the data and the missingness mechanisms. The missingness patterns are all the possible arrangements of missing and observed values in the data. In longitudinal studies, three types of missingness patterns are usually described. The first is a univariate pattern where only one variable has missing data. The second type is monotone missingness which occurs when, once a patient has a missing observation at one time-point, values for all subsequent time-points are also missing. The third general pattern of missingness is arbitrary (or intermittent) missingness in which there is no specific structure for the measurements and patients with missing values. These patterns relate to the different ways in which missing data occurs in longitudinal studies. If patients are lost to follow-up, their exposure, covariates and outcome are completely missing at a specific time-point and onwards, corresponding to monotone missingness. Sparse follow-up is another context in which exposure, covariates and outcome can be fully missing, but unlike loss-to-follow-up, this missingness pattern is not monotone. In this paper, we focus on this third situation, as there is little guidance in the literature on the best statistical method to tackle this problem. This is also the most common pattern found in routinely collected data where data are not collected for a research purpose, where the quality of recording may vary from one visit to another.

Little and Rubin's classification is often used to classify the missing data as being (i) missing completely at random (MCAR) when the probability of data being missing does not depend on the observed or unobserved data, (ii) missing at random (MAR) if the probability of data being missing does not depend on the unobserved data, conditional on the observed data or (iii) missing not at random (MNAR) if the probability of data being missing depends on the unobserved data, even after conditioning on the observed data (22). The next section presents five methods for addressing missing confounder values in MSMs and relates their validity to the underlying missingness mechanism.

### 3.2. Missing data methods

#### 3.2.1. Complete case analysis

Complete case (CC) analysis is a straightforward and widely used approach to handle missing data in MSMs. In CC analysis, the parameters of the MSM are estimated from the sub-sample of patients with a complete record for all the variables and for all time-points (17). The first limitation of this approach is a loss in sample size, which may affect the precision of the estimates. The second, – and more problematic – limitation, is the risk of invalid inference following its use. In standard regression modelling, CC analysis can allow the estimation of unbiased conditional estimates of the treatment effect provided that the probability of being a complete case is independent of the outcome (23). The validity of CC analysis to estimate marginal effects relies on the stronger assumption that missing data are MCAR (24). This assumption is rarely plausible, and its violations can usually be demonstrated using the observed data.

#### 3.2.2. Multiple imputation

The general principle of MI is to generate multiple sets of plausible values for the missing observations by drawing from the posterior predictive distribution of these variables given the observed data. This leads to the generation of $M$ complete datasets, which are then analysed independently to produce $M$ estimates of the regression coefficients of the MSM. These estimates are then combined across the $M$ imputed datasets following Rubin's rules to obtain an overall estimate and its variance, as described in Appendix A2.

MI can provide valid inference if the data are MAR, regardless of the role of variables predictive of missingness in the analysis model (i.e. treatment, outcome or confounder). Unfortunately, the extensions of MI to accommodate data under MNAR mechanisms are limited (25). In addition, a correct specification of the imputation model is key to obtain valid inference after MI of the covariates. Critically, even when MI is used to handle missing values in confounder data, the outcome must be included in the imputation model in order to preserve the existing relationships between the variables and the outcome (26). Furthermore, in longitudinal studies, it has been recommended to include all the covariate measurements in the model as well as the full treatment history. In doing so, several measurements of the same variable are seen as independent variables, preserving the correlation between subsequent measurements. However, convergence problems may arise because of overfitting when the number of time-points is large in comparison to the total sample size (27). Providing that factors associated with missingness are measured and the imputation model is correctly specified, MI may lead to unbiased estimates of the treatment effect (17) estimated from MSMs.

### 3.2.3. The missingness pattern approach

The missingness pattern approach (MPA) is a method proposed to handle missing data in the specific context of propensity-score analysis, including IPTW (28,29). This approach involves splitting the analysis sample into subsamples based on their missingness pattern. Then, the weight models are estimated separately within each pattern using only the variables fully observed in that pattern. When using IPTW with a single time-point, the MPA can lead to the estimation of unbiased treatment effects under a set of assumptions which are not dependent on the missingness mechanism itself but rather on the existing relationships between the true (but unobserved) values of the confounders and the treatment and outcome (20,30). The first assumption is the absence of unmeasured confounding within each missingness pattern (which is the conditional exchangeability assumption introduced in Section 2.1 within each pattern). In addition, the MPA requires that the partially observed covariate is no longer a confounder once missing (for more details, see (20)). Because of the peculiarity of these assumptions, the MPA may be valid under some MNAR mechanisms but biased under some MCAR mechanisms. To our knowledge, this approach has not been extended yet to time-varying IPTW for MSMs, but might exhibit good performances providing the aforementioned assumptions hold at each time-point.

### 3.2.4. The Last observation carried forward

Due to its simplicity, the last observation carried forward (LOCF) is another popular approach for handling missing data in studies with repeated measurements. LOCF is a form of single imputation, in which a missing covariate value is replaced by the most recent value recorded for that covariate for the same patient. The LOCF approach requires a complete measurement of the baseline data, because no prior information would be available to impute the first value. However, the main limitation of this approach is that its validity relies on the strong assumption that (i) either the true (but unobserved) distribution of the missing values at a given time point is exactly the same as the distribution of the observations used for the imputation (31) or (ii) when a value is missing, the treatment decision depends on the previous available measurement rather than the true (unobserved) one (32). In the first situation, LOCF is valid when the reason for missingness is the absence of change in values between two subsequent measurements. For instance, in routinely collected primary care data, a patient who is recorded as a non-smoker may not have smoking behaviour re-recorded unless they report a change (i.e. that they have started smoking). This is a very peculiar form of MNAR mechanisms where the associations between the missing confounder value and the previous observed value is deterministic. The second situation may also occur in routinely collected primary care data: general practitioners may base the treatment decision on the last available test results if the most recent results are not yet available. In the implementation of the LOCF approach, the uncertainty surrounding the imputation of missing values is not accounted for, leading to a potential underestimation of the variance of the treatment effect estimates, and therefore to an inflation of the type I error rate (33).

MI can account for the uncertainty in the imputation but standard implementations of MI make the MAR assumption, so do not provide a direct alternative to LOCF when data follow this specific missingness mechanism.

### 3.2.5. Inverse-probability-of-missingness weighting

In MSMs, Robins and Hernán (10) proposed the use of censoring weights to account for patients lost to follow-up. Complete cases are re-weighted by the inverse of their probability of remaining in the study. Loss to follow-up can be viewed as a missing data problem, and therefore, these weights can be accommodated to account for missing data in an approach called inverse-probability-of-missingness weighting (IPMW). The main difference is that loss to follow-up will generate monotone missing data, that is, if data is missing at a given time-point, it will be missing for all subsequent time-points, whereas in MSMs, missing data are often sporadically missing. IPMW is very similar to IPTW introduced in section 2. Whereas IPTW aims to account for confounding by balancing the characteristics of treated and untreated patients, IPMW aims to additionally balance the characteristics of complete cases and incomplete cases (34). It thus involves the estimation of two sets of weights: treatment weights and missingness weights (see Appendix A3 for more details). The overall weight (i.e. the product of the weights of being treated and being a complete case) simultaneously deals with confounding and missing data. This approach was initially introduced in MSMs in the presence of loss to follow-up (10), meaning that, in our context of arbitrary missingness patterns, patients are censored once they have a missing confounder value, even when subsequent values are measured. Therefore, IPMW does not make use of all the information available, which can make it an inefficient technique for small to moderate sample sizes. However, IPMW might be preferred over MI when patients with missing data tend to have missing values on many, rather than just one or two, variables, as it is often the case in observational studies (34).

The missingness mechanisms encountered in electronic health records are often different from typical mechanisms observed in trials or observational cohorts created for research. Some of the *ad hoc* missing data methods presented above could be particularly useful for MSM analyses in this context.

## 4. Methods

We performed a simulation study to (i) illustrate the impact on bias of violations of the assumptions required for each method to be valid, and the relative precision of these methods when the assumptions hold and (ii) highlight existing challenges in their implementation in practice. Data were simulated to mimic an observational study, looking at the effect of a time-varying binary treatment on a continuous outcome, in the presence of time-varying confounding. Figure 1 presents a causal diagram illustrating the association between simulated variables. We focused on four plausible types of missingness mechanisms, illustrated in Figure 2- 3:

- MCAR mechanism: missingness is not dependent on either observed or unobserved variables
- MAR mechanism: we consider 3 situations in which missing data depends on observed past treatment and confounder values (MAR|A,L), past treatment and confounder values and outcome (MAR|A,L,Y) or considering an association between missingness and the outcome introduced through the independent risk factor (MAR|A,L,V).
- "Constant" mechanism: confounder values are missing if they have remained constant since the last visit. This is one of the mechanisms under which LOCF is expected to perform well.
- "Differential" mechanism: the missingness mechanism itself is MAR, but missingness affects the subsequent association between the true value of the confounder, and the treatment. This is the mechanism implicitly assumed when using the MPA.

We compared the performance of CC analysis, LOCF, MPA, MI and IPMW to estimate the causal effect of the intervention at each time-point. The data-generating mechanisms, methods, estimands and performance measures for our simulations are presented in Appendix A4.

## 5. Results

Table 1 presents the scenarios for which each method can provide unbiased estimates of the treatment effects. The results of our simulation studies are presented as boxplots, showing the distribution of the absolute bias for each method. Full results are presented in Appendix (Tables A1 to A5). Note that even without missing data, if time-varying confounders are not accounted properly, the treatment effect estimates are strongly biased (Table A1).

### 5.1. Missingness completely at random

Whereas, CC, MI and IPMW lead to unbiased estimates at the three-time-points, the MPA estimates are biased at each time-point and LOCF estimates are biased at times 1 and 2 (Figure 3). The bias for the MPA arises from the direct associations existing between the confounders and the treatment allocation at subsequent time-points even among participants with missing covariate values. For LOCF, the bias arises because the missing values were generally different from the observed previous value because confounder values were affected by prior treatment. Among the unbiased approaches, MI had the best precision. The loss in efficiency for CC is explained by the loss in sample size (going from 10,000 to 4,096). For IPMW, this loss is explained both by the same loss in sample size but also because of additional random variability introduced by the estimation of two independent sets of weights.

### 5.2. Missingness at random

Except MI, which led to unbiased estimates at each time-point for the three MAR scenarios, the performance of the other analysis strategies relied on the variables that were predictive of missingness (Figure 4). When missingness depended on the values of past treatment assignment and confounders, IPMW estimates were unbiased at the three time-points. A small bias was observed for complete case analysis and larger biases were obtained when using LOCF and MPA, for similar reasons as in MCAR scenarios. When the outcome was directly related to missingness, the only unbiased approached was MI. The bias of the treatment effect for all the other approaches was very large at the three time-points. However, when an indirect association between the outcome and missingness existed through a measured risk factor, the IPMW led to unbiased estimates, but with a lower precision than MI.

### 5.3. Missingness on constant values

When missing confounder values were missing when they remained constant, only LOCF allowed the estimation of unbiased treatment effect estimates (Figure 3). In that situation, the bias was worse with the use of MI and IPMW than with CC. This is because both MI and IPMW uses the existing relationships between the confounders, treatment and outcome among the participants with fully observed data, but in this scenario, these relationships do not reflect the associations existing between the true (but missing) confounders values and the other variables.

### 5.4. Missingness affecting the subsequent covariate-treatment associations

In this last scenario, MPA was the only appropriate method to obtain unbiased treatment effect estimates, although the bias observed for the other approaches is quite small. As in the previous scenario, the associations between confounders and treatment among the complete cases cannot be used to make inferences about the relationship among participant with missing confounder values. Therefore, CC, MI and IPMW are biased. LOCF does not lead to valid estimates either, because missingness depends on past treatment and confounder values, which induces missingness whatever the previous measurement value was.

## 6. Illustrative example

We used the data from a prospective national cohort, using the research database of the "Observatoire Sommeil de la Fédération de Pneumologie" (OSFP). The OSFP registry is a standardized web-based report, administered by the French Federation of Pulmonology and containing anonymized longitudinal data from patients with sleep disorders (35). We aimed to estimate the causal effect of compliance in the use of continuous positive airway pressure (CPAP) device on sleepiness symptoms among patients diagnosed with obstructive sleep apnoea. This device is a pump which delivers a continuous supply of air through a mask patients use overnight. In the OSFP registries, the number of recorded visits per patient varies, so for simplicity, we focused on patients who had a visit within 3 months, 6 months and 1 year following the initiation of CPAP treatment. The observance was determined as an average use of CPAP >4hours per night within each time interval. The outcome was a continuous sleepiness score measured during the last visit using the Epworth Sleepiness Scale. Age, sex, body mass index (BMI), urination frequency per night and presence of depression were considered as potential confounders in this study. BMI, frequency of urination and depression were recorded at each visit but were partially observed. Patients' characteristics are described in Appendix (Table A6).

Out of 1169 included patients, only 266 (22.8%) patients had a complete record. Data is not MCAR since significant associations were observed between all the potential confounders and the probability of having a complete record. However, a MAR mechanism is plausible in this setting. The MPA is not a suitable method for this analysis because of missing data on the exposure that can be handled with the other approaches (CC, LOCF, MI and IPMW). These approaches were implemented in the same way as in our simulation study. 95% confidence intervals were computed using non-parametric bootstrap.

The results are presented in Figure 5 and in Table A7. Overall, these results show no effect of CPAP compliance on sleepiness. As expected because of the relatively small sample size, IPMW led to very wide 95% confidence intervals. The 95% CI for LOCF is narrower, but it does not account for the uncertainty around the imputed values. Furthermore, the assumption underlying the validity of LOCF is unlikely to hold in this context. As expected, the approaches gave similar results because confounding was not very strong (Table A6). However, it illustrates the inefficiency of IPMW and the limitations of the MPA approach when there are missing data on the exposure.

## 7. Discussion

In this paper we presented five methods to handle missing values in partially observed time-varying covariates in MSMs and identified situations in which they are appropriate to estimate unbiased causal treatment effects. We also illustrated how to implement these approaches in practice, using data from the OSPF registry. We showed that, for the estimation of causal effects, CC analysis is often biased, unless data are missing completely at random. The validity of this assumption cannot be tested from the data (36) but violations of these assumptions can be detected by looking at associations between the probability of being a complete case and the variables available in the dataset, including the treatment and the outcome. While the missing completely at random assumption is rarely plausible in practice, we also showed that when missing values are missing at random given treatment history and confounder values, the bias of the CC estimates is usually small.

LOCF, which is also widely used to handle missing data on time-varying variables leads to biased estimations of the treatment effects, unless missing values are in truth missing because they remained constant over time or when the previous measurement is used to adapt treatment (rather than the true - but missing - measurement). This is a particular case of MNAR mechanism. This assumption may hold in routinely collected data. For instance, GPs might not record a patient's weight during a visit if it has not changed since the previous consultation. This assumption cannot be tested from the data but the plausibility of the assumption can be assessed using expert opinion, building on what has been proposed in randomised trials (37). Moreover, when using LOCF, the uncertainty around the single imputation of missing values is not accounted for (24). Although this is not an issue for categorical variables, this is problematic for continuous confounders where imputing exactly the previous measurement leads to inappropriate certainty, which might affect the validity of the inference made with too narrow confidence intervals.

As expected, MI led to unbiased estimates of the treatment effect when data were MCAR or MAR. When implementing MI, the outcome must be included in the imputation model and the treatment effect estimated in each imputed dataset and combined using Rubin's rules, as recommended in simpler settings with a single time-point (38). In our simulation study and illustrative example, treatment and covariate values for all the three time-points were included in the imputation model. With an increasing number of time-points, issues of overfitting may arise. Two-fold multiple imputation has been proposed to circumvent this problem (39). Instead of using all the time-blocks in the imputation model, only the current and adjacent times are used in addition to the outcome, and therefore fewer parameters are to be estimated in the imputation model. This approach showed good performance in the analysis of longitudinal electronic health records (27) even when the partially observed variable has a non-linear association with time (40).

MPA and IPMW have never been investigated in the context of MSMs. The MPA is unbiased when either the association between the partially observed confounder and the outcome or between the partially observed confounder and the subsequent treatment disappears among patients with a missing value for that confounder. Hence, the validity of this approach does not depend on the missingness mechanism but instead the relationship between covariates, treatment and outcome among patients with missing data. One situation in which the MPA is thought to be useful is in the analysis of routinely collected data in which treatment prescription depends on biological test results. If these results are missing (i.e. the test is not done for that patient), physicians cannot use the test value in their treatment decision, which removes the association between the true value of biological parameter (which is unknown) and the subsequent treatment. The MPA is straightforward to implement but issues may arise when there are many missingness patterns. Furthermore, unlike MI, the MPA cannot accommodate missing data on the treatment and the outcome, which might limit its applicability. IPMW, on the other hand, can handle missing data on any of the covariates, treatment or outcome, as it includes only the complete records in the analysis, regardless which variables (outcome, treatment or covariates) have missing values. IPMW leads to unbiased estimates when data is MCAR and MAR, unless missingness is directly affected by the outcome, because the outcome might be associated with treatment and confounder values at later time-points, which are not accounted for in the missingness model to compute the weights. IPMW is also unbiased in scenarios where the MPA is unbiased as well because patients are censored at the first missing data, and therefore, no use is made of the information measured at later time-points. However, IPMW is somewhat inefficient. This is explained firstly by a loss in sample size loss since only complete cases are analysed (34). Secondly, IPMW involves the multiplication of two weights (the treatment weight and the missingness weight) that are both estimated with uncertainty, leading to highly variable treatment effect estimates. However, the sample size in the simulation study and the illustrative example was moderate, so we recommend the use of IPMW in very large datasets, a situation in which MI would be highly computationally intensive. Furthermore, a limitation in the current implementation of IPMW is that missing data were considered as being monotone, that is, patents were excluded from the analysis even when some (or all) the information was available at the end of follow-up, including the outcome.

Recent developments on inverse weighting for missing data have included an extension to non-monotone missingness patterns (41), but it remains unclear how it could be transposed to MSMs.

It is clear that there is no single missing data method able to simultaneously handle different types of missingness mechanisms. However, in practice, missing values can occur on several variables according to different mechanisms. In such situations, it is crucial to understand the reasons for missingness to identify groups of variables with similar missingness mechanisms that could be handled altogether. For instance, in routinely collected data, some variables might not have been updated because their values remained unchanged, and some variables might be missing at random. A pragmatic approach would be to first use the LOCF on the first group of variables, and then multiply impute the variables from the second group. A more principled combination of methods has been proposed in simpler settings. Qu and Lipkovich (42) combined the MPA and MI for propensity score analysis with a single time point. Seaman and White proposed to combine MI and IPMW (43) but further investigation is needed before implementing these methods in the context of MSMs.

This paper has several limitations. First, we focused on a continuous outcome measured at the end of follow-up. However, our conclusions can be applied more broadly to binary and time-to-event outcomes, since the validity of CC, LOCF, MI, MPA and IPMW relies on the missingness mechanism which is independent of the nature of the outcome. Similar conclusions would also hold had the outcome been measured at each time-point. With repeated outcomes, however, intra-subject correlation must be accounting for in the outcome model, using for instance a sandwich variance estimator in the context of generalized estimated equations (44). Second, we considered a limited number of partially observed confounders in our simulation study. With a larger number of variables, the MPA approach might reach its limits in case of sparse data in some patterns of missingness. Moreover, in the presence of numerous variables with potential interactions and non-linear effects, the functional form of the weight model might be harder to specify parametrically, and the obtained weights can be very unstable. This can be improved using the robust methods proposed by Imai and Ratkovic (45). Alternatively, statistical learning methods might be a promising alternative to pooled logistic regression in such settings (46). Third, we acknowledge that the standard error of the treatment effect estimates in the simulation study did not account for the uncertainty in the estimations of the weights, which lead to 95% confidence intervals that were too large. A variance estimator accounting for this uncertainty has been proposed in the context of propensity score weighting (47) but has yet to be extended to the MSM context. Non-parametric bootstrap was used in our illustrative example but was unfortunately too computationally demanding for use in our simulation study. However, this impacts all methods equally and therefore should not affect the comparison of the results.

In conclusion, the choice of the appropriate method(s) to handle partially observed confounders when using MSMs must rely on a careful consideration of the reasons for missingness and whether missingness modifies the existing relationships among observed data. Although MI outperforms the other approaches when data are missing at random, we presented two scenarios, encountered in routinely collected data, where MI leads to biased estimates of the treatment effect estimates but LOCF and the MPA might be suitable alternatives. We end by noting that any analysis using data with some missing values inevitably relies on some assumptions about the missingness mechanisms or missingnesss patterns, and that these assumptions are often not made explicit. We therefore encourage researchers to clearly describe the assumptions under which their primary analysis is valid, and to perform sensitivity analyses to assess robustness of their results to departures from these postulated missingness mechanisms.

Table 1. Summary of findings

| Mechanism | | Complete cases | Last observation carried forward | Missing pattern approach | Multiple imputation | Inverse probability of missingness weighting |
|---|---|---|---|---|---|---|
| MCAR | | ✓ | ✗ | ✗ | ✓ | ✓ |
| MAR | | | | | | |
| | A,L | ✗ | ✗ | ✗ | ✓ | ✓ |
| | A,L,Y | ✗ | ✗ | ✗ | ✓ | ✗ |
| | A,L,V | ✗ | ✗ | ✗ | ✓ | ✓ |
| Constant | | ✗ | ✗ | ✓ | ✗ | ✓ |
| Differential | | ✗ | ✓ | ✗ | ✗ | ✗ |

Red crosses indicate biased methods, green ticks indicates unbiased methods

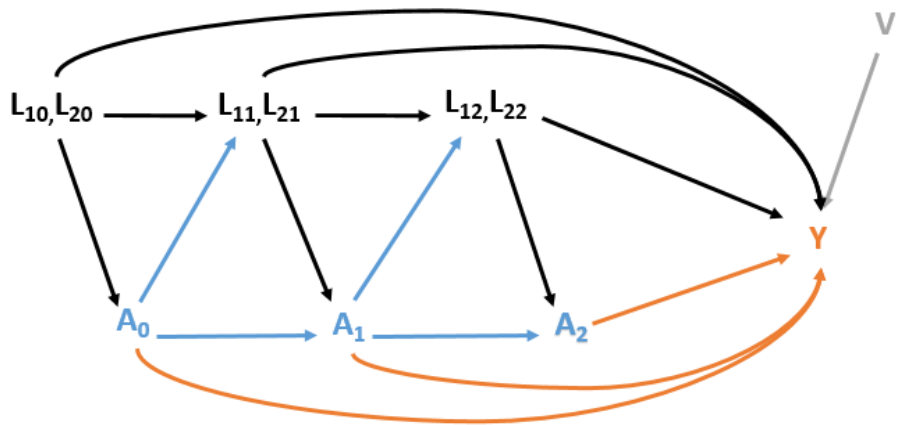

Figure 1. DAG representing the associations between simulated variables.

$L_1, L_2$ are the two time-varying confounders, A is the time-varying treatment, Y is the continuous outcome, V is a time-invariant independent risk factor.

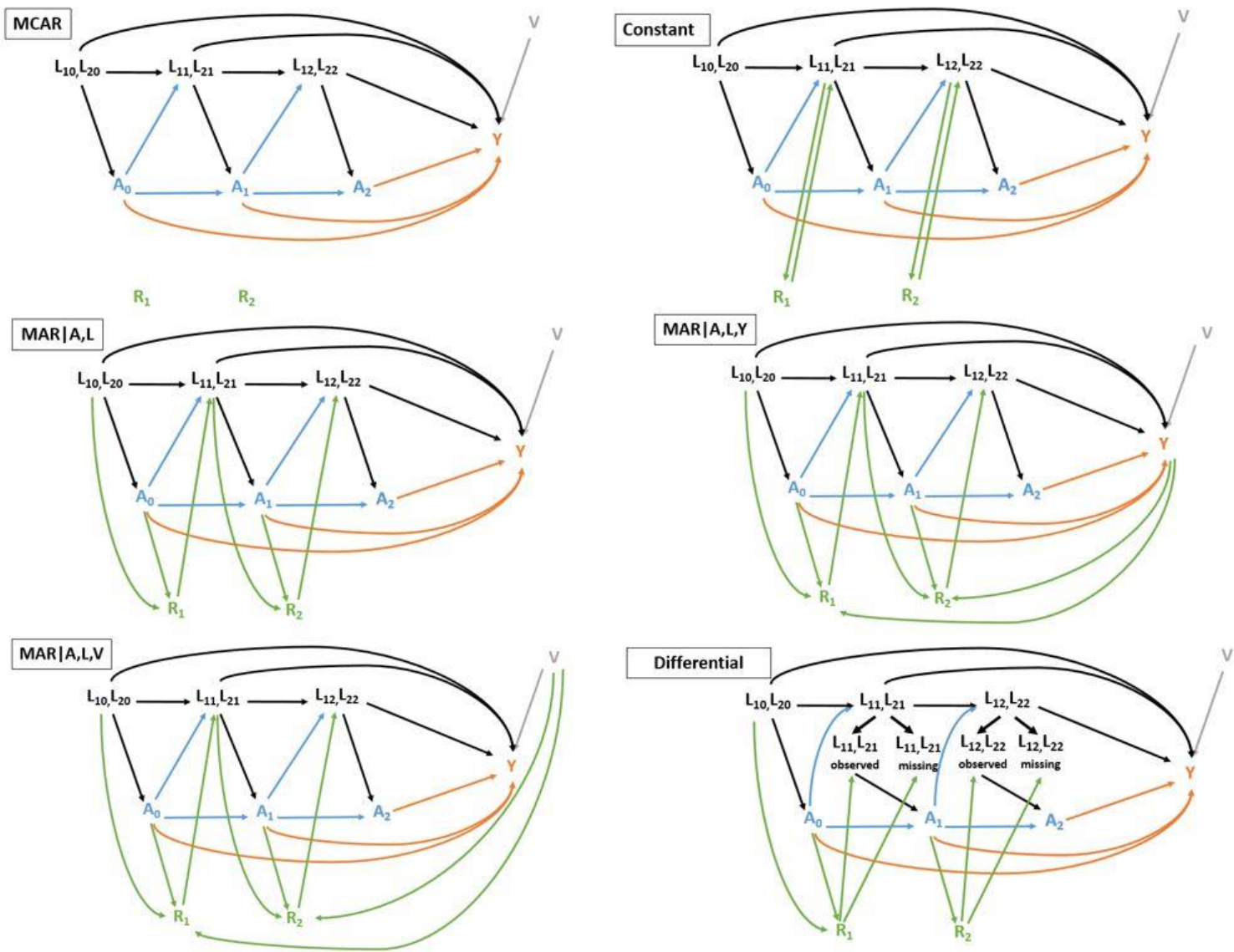

Figure 2. DAGs representing the scenarios in which missing values are missing completely at random (MCAR), missing at random (MAR) given the confounders and the treatment (MAR|A,L), given the confounders, the treatment and the outcome (MAR|A,L,Y) and given the confounders, the treatment and the independent risk factor (MAR|A,L,V). A DAG the scenario in which missingness occurs when the confounder value has remained unchanged since the previous measurement (Constant). The double arrow indicates that if there is no change in the confounder values from the previous time-point. A DAG also presents the scenario in which missingness depends on the treatment and confounders (Differential) but the association between the missing value and the subsequent treatment allocation does no longer exist. The confounder only contributes to the treatment allocation decision when observed.

$L_1, L_2$ are the two time-varying confounders, A is the time-varying treatment, Y is the outcome, V is an independent risk factor and $R_1$ and $R_2$ are the missingness indicators at k=1 and k=2, respectively.

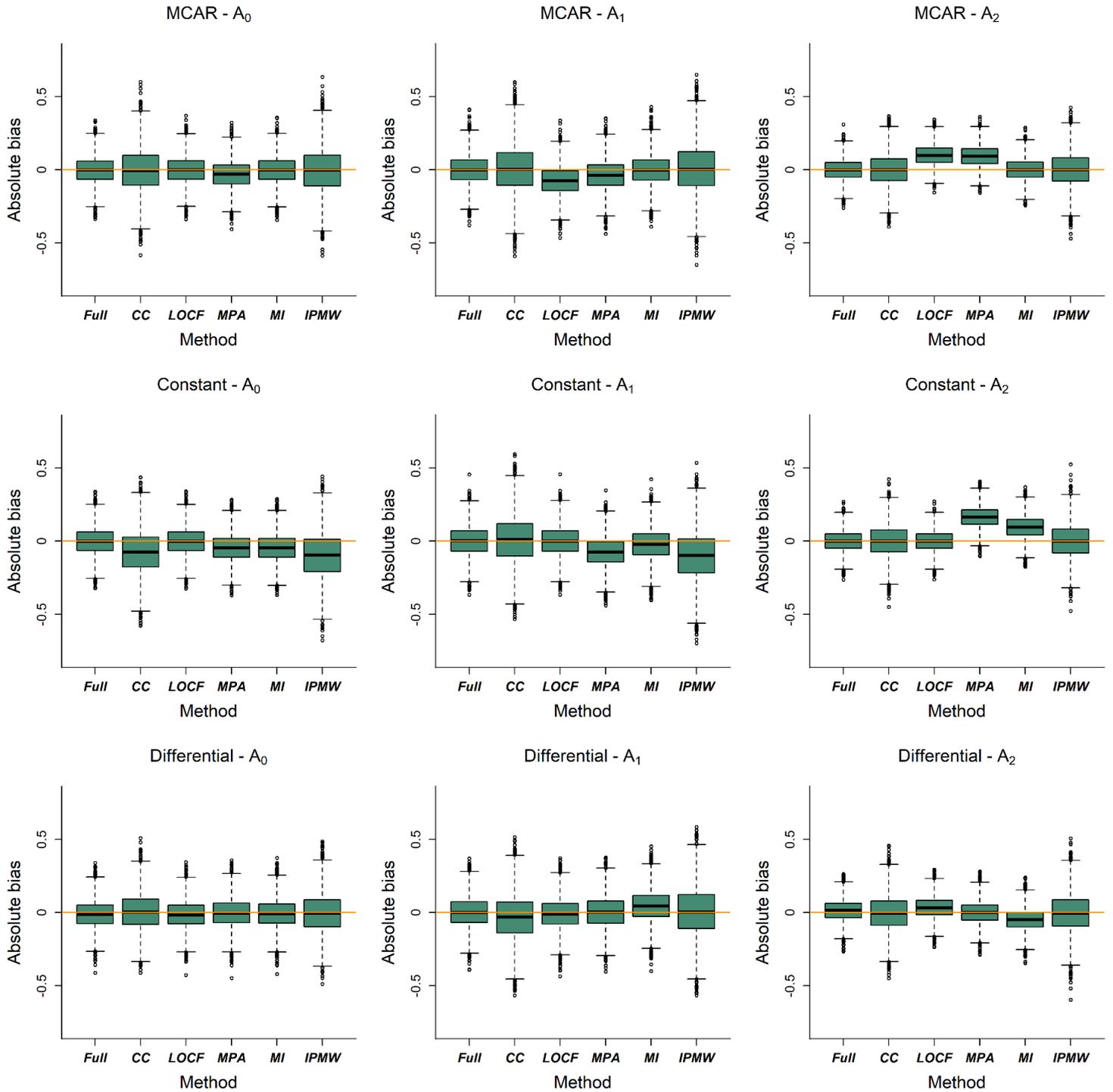

Figure 3. Absolute bias of the treatment effect estimate at k=0, k=1 and k=2 on full data and following the use of different missing data approach under the missing completely at random (MCAR), Constant and Differential missingness mechanisms, according . N=10000.

CC: complete cases; LOCF: last observation carried forward; MPA: missing pattern approach; MI: multiple imputation; IPMW: inverse probability of missingness weighting. For multiple imputation, 10 imputed datasets were generated.

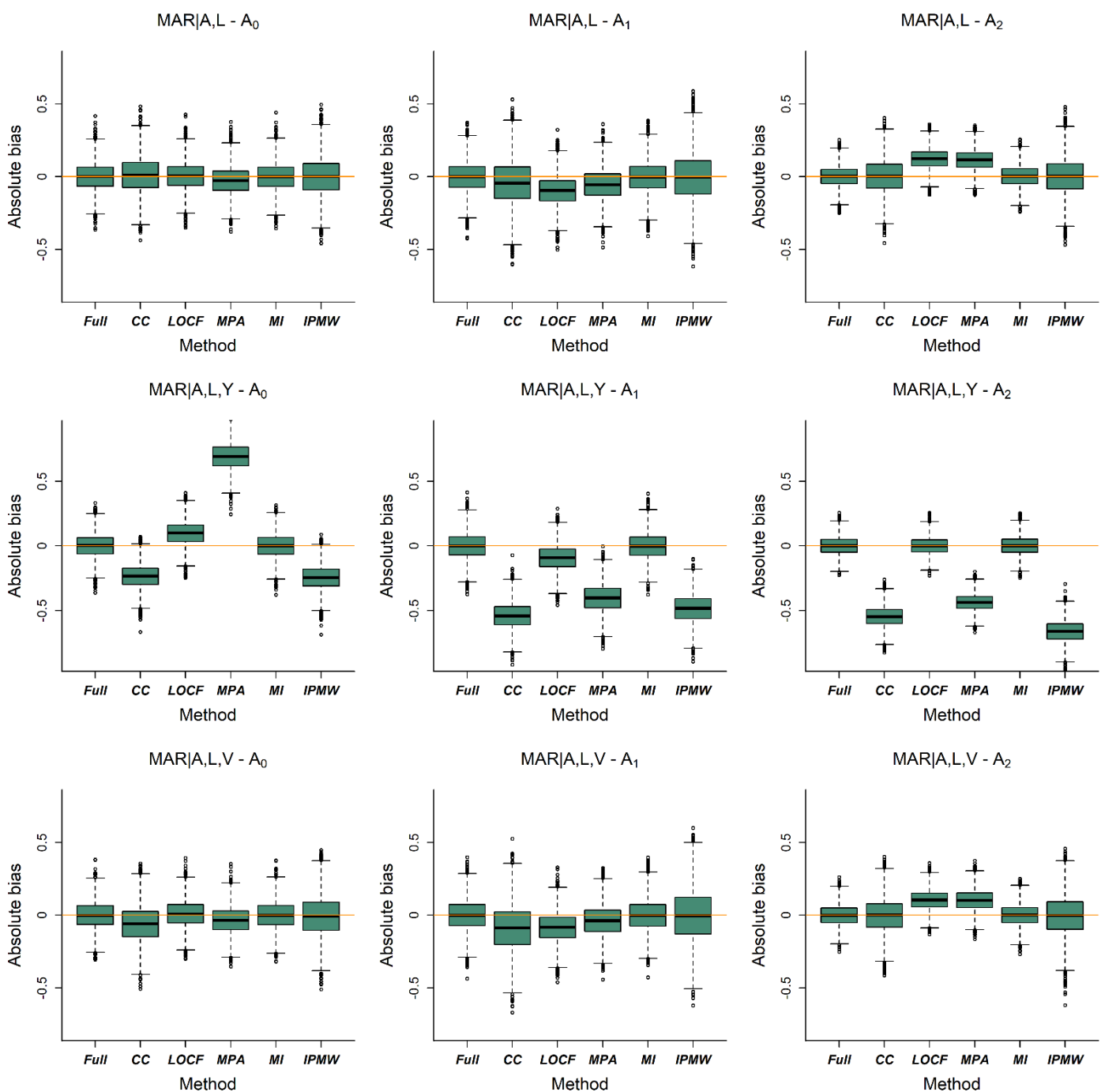

Figure 4. Absolute bias of the treatment effect estimate at k=0, k=1 and k=2 on full data and following the use of different missing data approach under 3 scenarios of data missing at random: missing at random given the covariates and the treatment (MAR|A,L), given the covariates, the treatment and the outcome (MAR|A,L,Y) and given the covariates, the treatment and the independent risk factor (MAR|A,L,V). N=10000.

CC: complete cases; LOCF: last observation carried forward; MPA: missing pattern approach; MI: multiple imputation; IPMW: inverse probability of missingness weighting. For multiple imputation, 10 imputed datasets were generated.

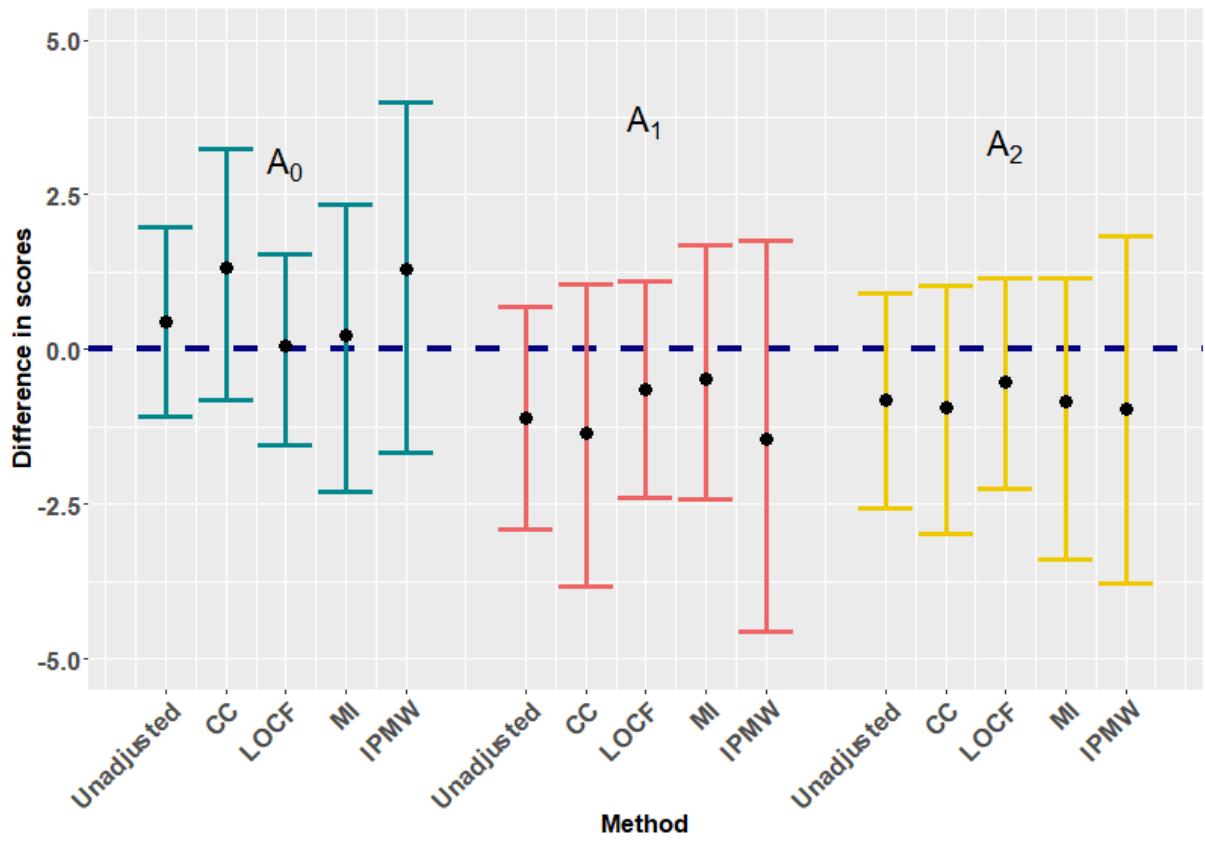

Figure 5: Results of the illustrative example.

# APPENDICES

## A1- The potential outcome framework and causal inference in single-time studies

Causal inference is often formalised using the potential outcomes framework (1), initially developed in the context of randomised experiments. In a simple setting in which one is interested in estimating the causal effect of a binary treatment A on an outcome Y measured at a single time-point, the causal treatment effect is defined as the difference between the two potential outcomes $Y^{A=0}$ and $Y^{A=1}$. These are the outcomes that would have been observed had the population been treated versus not.

Three assumptions are usually made to consistently estimate the causal effect of a treatment: (i) no interference, (ii) consistency, and (iii) conditional exchangeability. Assumption (i) means that the potential outcome values for a given patient are not affected by the treatment values of other patients. Assumption (ii) implies that a patient has only one possible potential outcome value for each treatment under study and assumption (iii) ensures that there are no unmeasured confounders. If these assumptions hold, a multivariable regression model may allow the estimation of an unbiased conditional causal treatment effect, provided the model is correctly specified. More advanced statistical techniques, such as inverse-probability-of-treatment weighting (IPTW) may be used to estimate marginal causal effects. The validity of IPTW relies on the additional assumption of positivity (2), meaning that every patient has a non-null probability of receiving either treatment. IPTW aims to create a pseudo-population in which the treatment is independent of the confounders, analogous to the situation achieved by randomising the treatment in a trial, by weighting the patients by the inverse of their probability of receiving the treatment they actually received. This probability is derived from the propensity score (3) and can easily be estimated from the data using a logistic regression model. Both multivariable regression and IPTW are valid approaches to estimate causal effects (conditional and marginal, respectively) in a simple setting of a single time-point. However, challenges arise when treatment and confounder values vary over time.

## A2- Estimating the weights of a marginal structural model

Suppose there are K+1 measurement occasions, from k=0 (baseline) to the study end (k=K). Let A be a binary treatment, with $A_k$ representing its level at time k (k=0,…,K) and $\bar{A}_k$ the treatment history until time *k*. $L_k$ is a vector of confounders measured at time *k*. We assume that once a patient receives the treatment (A=1), they remain under treatment until the end of follow-up (k=K) where the outcome Y is measured. We are interested in the estimation of $E(Y^{\bar{a}})$, the value of the outcome that would have been observed had all subjects received treatment history $\bar{a}$. or contrasts of these quantities. For example, in our simulation study (Section 4 and 5), we estimate the effect of treatment initiation at each time-point.

To estimate these quantities, the weights of the MSM are defined as the patient's probability of receiving their own treatment history given their health history. The stabilized weight for patient *i* (i=1,…n) is defined as:

$$sw_i = \frac{\prod_{k=0}^{K} P(A_k=a_{ki}|\bar{A}_{k-1}=\bar{a}_{(k-1)i})}{\prod_{k=0}^{K} P(A_k=a_{ki}|\bar{A}_{k-1}=\bar{a}_{(k-1)i}, \bar{L}_k=\bar{l}_{ki})}, \qquad (1)$$

where *k=1,…,K* is the *k*th time-point, $A_k$ the treatment received at time k, $\acute{A}_k$ the treatment history until time *k*, $L_k$ the covariates values at time *k* and $\acute{L}_k$ the covariates history until time *k*. Stabilized weights are preferred to standard weights (i.e. weights computed with 1 at the numerator), because they reduce the variability in treatment probabilities that may happen when some covariates are strong predictors of treatment. In practice, both the numerator and denominator can be estimated using pooled logistic regression, not including and including confounder history, respectively. In pooled logistic regression, each person-time interval is considered as an observation. This pooled logistic regression model must include the confounders and their relevant interactions to ensure the distributions of confounders are balanced between treatment groups in the weighted pseudo-population at each time-point.

**A3- Rubin's rules to combine treatment effect estimates after multiple imputation**

Multiple imputation of partially observed datasets leads to the generation of *M* complete datasets which are then analysed independently to produce *M* estimates $\hat{\theta}_k$, (*k = 1,..., M*) of *θ*, the vector of the parameters of interest (the regression coefficients for $A_0,...,A_t$ in our setting) and their associated variance matrix $\mathbf{W}_k$. Then, $\hat{\theta}_k$ and $\mathbf{W}_k$, are combined across the M imputed datasets following Rubin's rules in order to obtain an overall estimate and its variance, accounting for the noise introduced by the random elements of the imputation. These rules state that the overall estimate $\hat{\theta}_{MI}$ and its estimated variance $\widehat{Var}(\hat{\theta}_{MI})$, are (4):

$$\hat{\boldsymbol{\theta}}_{MI} = \frac{1}{M}\sum_{k=1}^{M}\hat{\boldsymbol{\theta}}_k, \qquad \widehat{Var}(\hat{\boldsymbol{\theta}}_{MI}) = \boldsymbol{W} + \left(1 + \frac{1}{M}\right)\boldsymbol{B},$$

where **W** is the within-imputation covariance matrix, reflecting the variability in the estimates of the parameters of interest within each imputed dataset and **B** is the between-imputation covariance matrix, reflecting how missing data impacts the variability in the estimates. These two variance components are estimated as:

$$\boldsymbol{W} = \frac{1}{M}\sum_{k=1}^{M}\boldsymbol{W}_k, \qquad \boldsymbol{B} = \frac{1}{M-1}\sum_{k=1}^{M}(\hat{\boldsymbol{\theta}}_k - \hat{\boldsymbol{\theta}}_{MI})^2.$$

Because the confounders only enter the first stage of the MSM process, it may be tempting to combine the weights across imputed datasets, rather than combining the final treatment effect estimates. However, this approach is known to provide biased estimates (5).

**A4- Estimation of the missingness weights**

IPMW is very similar to IPTW introduced in section 2. Thus, weights are estimated using, for instance, equation (1) (Appendix A1), but using $R_k$, the missingness indicator at time *k* ($R_k$=1 for complete cases and $R_k$=0 if at least one confounder has a missing value at time *k*). Under MCAR and MAR mechanisms, IPMW can allow the estimation of unbiased treatment effect estimates if the weight model is correctly specified (i.e. includes all the predictors of missingness in the correct functional form). The individual's missingness weights at time *t* are the probabilities of being a complete case up to that time.

**A5- Design of the simulation study**

*Aims*

We performed a simulation study (i) to illustrate the impact on bias of violations of the assumptions required for each method to be valid, and the relative precision of these methods when the assumptions hold and (ii) to highlight existing challenges in their implementation in practice.

*Data-generating mechanisms*

Data were simulated to mimic an observational study, looking at the effect of a time-varying binary treatment on a continuous outcome, in the presence of time-varying confounding. Figure 1 presents a causal diagram illustrating the association between simulated variables. There are three measurement occasions, indexed by k=0,1,2. The continuous outcome was measured at the end of follow-up (*k*=2). Two time-varying confounders (one binary, L1, and one continuous, L2) were measured at baseline (*k*=0) and at two subsequent visits (*k*=1 and *k*=2) and one independent risk factor was measured at baseline only. The first treatment prescription was dependent on the baseline value of these 3 variables. The treatment prescription was then updated based on the current values of the two time-varying confounders and treatment history, creating a treatment-confounder feedback loop. For simplicity, we assumed that after treatment initiation, patients remained under treatment. The two time-varying confounders were fully observed at baseline, but were partially observed at the two subsequent visits. Patients could have either none, one or the two confounder values missing at visits 2 and 3. Missingness was arbitrary (i.e non-monotone); a patient could have a missing value at *k*=1 but a measured one at *k*=2. For a full understanding of the data-generating mechanisms, the R code to generate the data is provided as a supplementary file.

Under each scenario, around 40% of missing date were present at time 2 and 3, leading to a proportion of complete cases for all time-points of about 40%.

*Methods*

Treatment effects estimates were obtained via MSM with IPTW as described in Section 3.2.5. At time-point *k* (*k*=0,1,2), inverse-probability-of-treatment weights were estimated using a logistic regression model with the treatment received at time t as the outcome and the two confounders $L_1$ and $L_2$ at that time-point as predictors. Only participants untreated at the previous time-point contributed to the estimation of the weights since participants remain under treatment once they have initiated it. Consequently, individual weights stayed constant after initiation. The final weights used for the analysis of the primary outcome were the product of the weights up to the time of initiation or up to time 2, whichever occurred first. The outcome model was a weighted linear regression model including binary treatment indicators at the three time-points.

We compared the following methods to handle missing data:

- CC analysis: the parameters of the MSM are estimated using complete cases only, that is patients with a measurement available for each of the two time-varying confounders at each of the three time points.
- LOCF: when a patient has a missing confounder value at time *k*, it is imputed by the most recent value observed for that confounder. These singly-imputed values are then treated as if they were the observed values; thus, the treatment effect estimates are obtained using a MSM from this single imputed dataset.
- MI: missing confounders values are imputed 10 times using multiple imputation by chained equations. The imputation model includes the treatment indicator and confounder values at the three time-points, as well as the independent risk factor and the outcome. The MSM is fitted in each imputed dataset and the 10 treatment effect estimates are pooled using Rubin's rules.
- MPA: at times *k*=2 and *k*=3, missing data could occur on $L_1$ or $L_2$, leading to 4 missingness patterns at each of these time-points. A different weight model is used within each pattern, including only the variables fully observed in that pattern. Therefore, weights are estimated for every patient based on available information, and the analysis model is fitted on the entire original sample.
- IPMW: the weights for being a complete case are estimated at each time-point. At baseline (k=0), there is no missing data, so the weight is 1 for everyone. Then, for *k*=2 and *k*=3, the weights are estimated as follows: first, a logistic regression model is fitted separately for each time-point. The outcome is the binary missingness indicator at that time-point, and the predictors are the independent risk factor, and the treatment and confounders history. The weights at the 3 time-points are then multiplied together to obtain an overall weight for being a complete case, itself multiplied by the overall treatment weight (estimated as in the CC analysis). In the second stage (the analysis model to obtain the MSM parameters), complete cases were re-weighted using the pooled weights (instead of the treatment weights).

*Estimands*

The three estimands of interest were the causal mean differences in the continuous outcome between treated and untreated participants at time $k$: $\hat{\theta}_0$ is the direct effect of the treatment $A_0$, unmediated by $A_1$ and $A_2$. Similarly, $\hat{\theta}_1$ is the direct effect of $A_1$, unmediated by $A_2$ and $\hat{\theta}_2$ is the direct effect of $A_2$.

*Sample size and number of simulations*

The sample size was 10,000 in each generated dataset and 5000 replications were used.

*Performance measures*

The performance of the five methods was assessed using the following measure:

- Bias of the treatment effect estimate at each time point $k$ ($k=0,1,2$): $B_k = E(\hat{\theta}_k) - \theta_k$ estimated as: $\frac{1}{5000}\sum_1^{5000} \hat{\theta}_{ki} - \theta_k$, where $\theta_k$ is the true value of the treatment effect at time $k$ and $\hat{\theta}_{ki}$ the estimate of the treatment effect in the $i^{th}$ simulated dataset (i=1,…,5000).
- The empirical standard error of the treatment effect estimate at time $k$: $SE_k = \sqrt{Var(\hat{\theta})}$, estimated from the data as: $\sqrt{\frac{1}{4999}\sum_{i=1}^{5000}(\hat{\theta}_{ki} - \bar{\theta}_k)^2}$, where $\bar{\theta}_k$ is the average treatment effect at time $k$ across the 5000 simulated datasets.
- The coverage rate, defined as the proportion of 95% confidence intervals containing the true value of the treatment effect.

The Monte-Carlo standard errors for these measures were also computed, as suggested by Morris *et al.* (6). All simulations were performed using R 3.5.1. The *mice* package (7) was used for multiple imputation using chained equations and the package *Survey* was used to conduct the weighted linear regression analysis (8).

**TABLES**

Table A1. Absolute bias and coverage rate (%) before the generation of missing values, for unadjusted, covariate adjusted and MSM analyses.

| Method | A0 | | A1 | | A2 | |
|---|---|---|---|---|---|---|
| | Bias | Coverage | Bias | Coverage | Bias | Coverage |
| Unadjusted | 0.331 | 9.6 | -0.187 | 61.6 | 0.462 | 0.0 |
| Multivariable adjustment | -0.229 | 29.1 | -0.592 | 0.0 | 0.003 | 94.8 |
| MSM | 0.000 | 97.2 | 0.001 | 97.4 | 0.000 | 97.9 |

n=10000. The unadjusted analysis does not account for covariates other than treatment history. With multivariable adjustment, confounders at each time point are included in the outcome model. MSM: marginal structural model. 5000 simulations were performed. The maximum Monte-Carlo standard error was 0.002 for the bias and 0.005 for the coverage rate.

Table A2. Absolute bias and coverage rate (%) for the 5 methods to handle missing data in each scenario considered at time 0.

| Scenario | CC | | LOCF | | MPA | | MI | | IPMW | |
|---|---|---|---|---|---|---|---|---|---|---|
| | Bias | Coverage | Bias | Coverage | Bias | Coverage | Bias | Coverage | Bias | Coverage |
| MCAR | -0.003 | 97.6 | 0.000 | 97.5 | -0.030 | 96.3 | -0.001 | 97.3 | -0.004 | 96.4 |
| MAR\|AL | 0.010 | 97.2 | 0.004 | 97.2 | -0.029 | 96.3 | 0.000 | 97.0 | 0.000 | 96.6 |
| MAR\|ALY | -0.234 | 36.3 | 0.099 | 88.8 | 0.694 | 0.0 | 0.002 | 97.1 | -0.245 | 36.3 |
| MAR\|ALV | -0.060 | 95.5 | 0.009 | 97.7 | -0.034 | 96.5 | 0.000 | 97.7 | -0.008 | 96.4 |
| Constant | -0.074 | 95.7 | 0.000 | 97.1 | -0.046 | 95.5 | -0.045 | 95.8 | -0.098 | 93.2 |
| Differential | 0.006 | 97.3 | -0.013 | 97.3 | -0.002 | 97.0 | -0.007 | 97.3 | -0.005 | 96.9 |

CC: complete cases; LOCF: last observation carried forward; MPA: missing pattern approach; MI: multiple imputation; IPMW: inverse probability of missingness weighting. The initial sample size was n=1000. After introduction of missing data, the average number of complete cases varied between n=4096 and n=5698, depending on the scenario. For multiple imputation, 10 imputed datasets were generated. 5000 simulations were performed. The maximum Monte-Carlo standard error was 0.002 for the bias and 0.005 for the coverage rate.

Table A3. Absolute bias and coverage rate (%) for the 5 methods to handle missing data in each scenario considered at time 1.

| Scenario | CC | | LOCF | | MPA | | MI | | IPMW | |
|---|---|---|---|---|---|---|---|---|---|---|
| | Bias | Coverage | Bias | Coverage | Bias | Coverage | Bias | Coverage | Bias | Coverage |
| MCAR | 0.005 | 96.9 | -0.074 | 93.7 | -0.036 | 96.1 | -0.001 | 97.4 | 0.006 | 96.3 |
| MAR\|AL | -0.043 | 97.0 | -0.096 | 89.8 | -0.055 | 95.2 | -0.003 | 97.1 | -0.005 | 97.0 |
| MAR\|ALY | -0.540 | 0.1 | -0.093 | 92.6 | -0.404 | 25.4 | -0.002 | 98.1 | -0.485 | 1.9 |
| MAR\|ALV | -0.089 | 94.7 | -0.085 | 91.8 | -0.039 | 96.0 | -0.001 | 97.0 | -0.004 | 96.8 |
| Constant | 0.012 | 97.9 | 0.000 | 97.4 | -0.072 | 93.5 | -0.021 | 97.1 | -0.100 | 94.4 |
| Differential | -0.032 | 97.5 | -0.009 | 97.3 | 0.003 | 96.9 | 0.045 | 96.0 | 0.006 | 97.1 |

CC: complete cases; LOCF: last observation carried forward; MPA: missing pattern approach; MI: multiple imputation; IPMW: inverse probability of missingness weighting. The initial sample size was n=1000. After introduction of missing data, the average number of complete cases varied between n=4096 and n=5698, depending on the scenario. For multiple imputation, 10 imputed datasets were generated. 5000 simulations were performed. The maximum Monte-Carlo standard error was 0.002 for the bias and 0.005 for the coverage rate.

Table A4. Absolute bias and coverage rate (%) for the 5 methods to handle missing data in each scenario considered at time 2.

| Scenario | CC | | LOCF | | MPA | | MI | | IPMW | |
|---|---|---|---|---|---|---|---|---|---|---|
| | Bias | Coverage | Bias | Coverage | Bias | Coverage | Bias | Coverage | Bias | Coverage |
| MCAR | 0.000 | 98.1 | 0.100 | 80.2 | 0.094 | 82.8 | 0.002 | 97.9 | 0.000 | 97.0 |
| MAR\|AL | 0.002 | 98.1 | 0.122 | 71.3 | 0.115 | 75.6 | 0.004 | 97.7 | 0.002 | 97.2 |
| MAR\|ALY | -0.547 | 0.0 | 0.000 | 98.7 | -0.437 | 0.0 | 0.002 | 98.3 | -0.663 | 0.0 |
| MAR\|ALV | -0.002 | 97.9 | 0.104 | 79.3 | 0.103 | 79.7 | 0.001 | 97.8 | -0.003 | 96.8 |
| Constant | 0.001 | 98.1 | 0.001 | 97.8 | 0.165 | 49.9 | 0.095 | 83.8 | 0.001 | 97.6 |
| Differential | -0.004 | 97.9 | 0.034 | 96.2 | 0.001 | 96.8 | -0.048 | 94.9 | -0.003 | 96.9 |

CC: complete cases; LOCF: last observation carried forward; MPA: missing pattern approach; MI: multiple imputation; IPMW: inverse probability of missingness weighting. The initial sample size was n=1000. After introduction of missing data, the average number of complete cases varied between n=4096 and n=5698, depending on the scenario. For multiple imputation, 10 imputed datasets were generated. 5000 simulations were performed. The maximum Monte-Carlo standard error was 0.002 for the bias and 0.005 for the coverage rate.

Table A5- Mean squared error of the mean differences for the different methods.

| | A0 | | | | | | A1 | | | | | | A2 | | | | | |
|---|---|---|---|---|---|---|---|---|---|---|---|---|---|---|---|---|---|---|
| | Full | CC | LOCF | MPA | MI | IPMW | Full | CC | LOCF | MPA | MI | IPMW | Full | CC | LOCF | MPA | MI | IPMW |
| MCAR | 0.009 | 0.023 | 0.009 | 0.010 | 0.009 | 0.025 | 0.011 | 0.028 | 0.016 | 0.013 | 0.012 | 0.030 | 0.005 | 0.013 | 0.015 | 0.014 | 0.006 | 0.014 |
| MARAL | 0.009 | 0.016 | 0.009 | 0.010 | 0.010 | 0.019 | 0.011 | 0.027 | 0.020 | 0.015 | 0.012 | 0.028 | 0.005 | 0.014 | 0.020 | 0.019 | 0.006 | 0.016 |
| MARALY | 0.009 | 0.063 | 0.019 | 0.492 | 0.009 | 0.070 | 0.010 | 0.303 | 0.019 | 0.175 | 0.011 | 0.248 | 0.005 | 0.306 | 0.005 | 0.196 | 0.005 | 0.447 |
| MARALV | 0.009 | 0.020 | 0.009 | 0.010 | 0.009 | 0.020 | 0.011 | 0.034 | 0.018 | 0.013 | 0.012 | 0.033 | 0.005 | 0.014 | 0.016 | 0.016 | 0.006 | 0.020 |
| Constant | 0.009 | 0.028 | 0.009 | 0.011 | 0.011 | 0.035 | 0.011 | 0.026 | 0.011 | 0.016 | 0.012 | 0.040 | 0.005 | 0.012 | 0.005 | 0.033 | 0.015 | 0.014 |
| Differential | 0.009 | 0.017 | 0.010 | 0.010 | 0.010 | 0.019 | 0.011 | 0.026 | 0.012 | 0.013 | 0.014 | 0.029 | 0.006 | 0.015 | 0.007 | 0.006 | 0.008 | 0.018 |

Table A6. Description of the exposure groups at baseline (illustrative example)

|  | Non observant | Observant | Standardised mean difference (%) |
|---|---|---|---|
|  | n*=204 | n*=1084 |  |
| Female | 70 (34.3) | 335 (30.9) | 7.3 |
| Age (mean (SD)) | 56.02 (13.33) | 58.26 (12.97) | 17.1 |
| Sedentary | 29 (14.3) | 184 (17.0) | 7.5 |
| BMI (mean (SD)) | 32.12 (6.48) | 32.19 (6.34) | 1.1 |
| Depression score** (mean (SD)) | 4.55 (3.91) | 4.35 (3.87) | 5.2 |
| Frequency of urination>1 | 87 (72.5) | 405 (77.1) | 10.7 |

* 153 patients had missing exposure value; ** Depression score was calculated using the Pichot depression scale

Table A7. Estimated mean differences in sleepiness scores for the different analysis methods.

| Analysis method | n | Mean difference (95% confidence interval) | | |
|---|---|---|---|---|
|  |  | t=1 | t=2 | t=3 |
| Unadjusted | 897 | -0.684 [-1.582;-0.213] | 0.152 [-0.868;1.171] | -1.324 [-2.278;-0.370] |
| IPTW |  |  |  |  |
| *Complete cases* | 266 | -0.257 [-2.013;1.500] | 0.268 [-2.203;2.739] | -0.072 [-2.407;2.263] |
| *LOCF* | 521 | -0.537 [-2.001;0.926] | -0.125 [-2.012;1.762] | -0.373 [-1.834;1.089] |
| *Multiple imputation* | 1169 | -0.804 [-1.768;0.161] | -0.128 [-1.293;1.036] | -1.194 [-2.200;-0.188] |
| *IPMW* | 266 | -0.381 [-2.235;1.473] | 0.116 [-2.371; 2.601] | 0.050 [-2.223;2.323] |
| *MPA* |  | NA | NA | NA |